\documentclass[10pt]{iopart}
\usepackage{graphicx,iopams,upgreek}
\usepackage{graphicx}% Include figure files
\usepackage{bm}% bold math
\usepackage{verbatim}% bold math
\usepackage{overpic}
\usepackage[usenames,dvipsnames]{xcolor}
\usepackage[colorinlistoftodos]{todonotes}
\usepackage{siunitx}
\usepackage{soul}
\usepackage{bm,color,amssymb}
\usepackage{color,nameref}
\newcommand{\az}[1]{{\color{black}#1\color{black}}}
\newcommand{\azz}[1]{{\color{black}#1\color{black}}}

\usepackage{cite}
\begin{document}
\title[Driven spheres, spheroids and rods in explicitly modeled polymer solutions]{Driven spheres, ellipsoids and rods in explicitly modeled polymer solutions}
\author{Andreas Z\"{o}ttl$^{1,2,3}$ and Julia M.~Yeomans$^1$}
\address{$^1$ The Rudolf Peierls Centre for Theoretical Physics, University of Oxford,\newline   Clarendon Lab., Parks Rd., Oxford, OX1 3PU, United Kingdom}
\address{$^2$ Institut für Theoretische Physik, TU Wien, Wiedner Hauptstraße 8-10, A-1040 Wien, Austria}
%\address{$^2$ PMMH, UMR 7636 CNRS-ESPCI-Universit\'{e}s Pierre et Marie Curie and Denis Diderot, 10, rue Vauquelin, 75231 Paris Cedex 5, France}
\address{$^3$ Erwin Schr\"odinger Int.\ Institute for Mathematics and Physics, University of Vienna, Boltzmanngasse 9, 1090 Wien, Austria}
\ead{andreas.zoettl@physics.ox.ac.uk}

\begin{abstract}
  Understanding the transport of driven nano- and micro-particles in complex fluids is of relevance for many biological and technological applications. Here we perform hydrodynamic multiparticle collision dynamics simulations of spherical and elongated particles driven through polymeric fluids containing different concentrations of polymers.
\az{We  determine the mean particle velocities   which are larger than expected from Stokes law for all particle shapes and polymer densities. Furthermore we measure the fluid flow fields and local polymer density \azz{and polymer conformation} around the particles. 
 %We find that local polymer-depleted regions \ju{close to \sout{around}} the particles are \ju{responsible} for an apparent tangential slip velocity near the particle which accounts for the measured flow fields and transport velocities. They are compared to a simple  two-fluid model for the viscosities which match the simulated results well.}
%  We show that the conformations of polymers around the driven particles are key to understanding the transport velocities in the polymeric fluids.
We find that polymer-depleted regions close to the  particles are responsible for an apparent tangential slip velocity which accounts for the measured flow fields and transport velocities. A simple two-layer fluid model gives a good match to the simulation results.}
\end{abstract}
%\keywords{active colloids, collective motion}
\submitto{\JPCM}
\maketitle
\ioptwocol

\tableofcontents

\vspace{2ex}

%\ju{Acknowledgements seem out of place in the ToC, but OK if journal style}

\section{Introduction}
\label{sec:intro}

Individual colloidal particles at equilibrium undergo Brownian motion
in Newtonian fluids such as water. Since the pioneering works of
Einstein, Langevin and Smoluchowski, an excellent understanding of this
dynamics has been developed \cite{Frey2005}.  By contrast,
the random motion of colloids in complex fluids such as polymer
solutions and gels, and in biological fluids and cells, is by
far less well understood although considerable progress has been made in
this research field during the last decades, both from the experimental
and the theoretical perspective (see, for example
  Refs.~\cite{Amblard1996,Metzler2000,Szymanski2009,Guo2012,Hofling2013}).
Even more challenging is understanding colloidal motion 
in complex fluids driven out of equilibrium by, for example, external forces or
fluid flow \cite{Menzel2015}, or if the system is intrinsically out of equilibrium as in 
the case of active colloids \cite{Zoettl2016,Bechinger2016}.

Understanding the motion of colloidal particles through polymer
solutions and polymeric or filamentous networks is of
considerable importance in colloid science and for biomedical
applications.  For example, using colloidal particles to probe the physical properties
  of the crowded environment of living cells 
  has been used to distinguish healthy cells from cancer cells
  \cite{Norregard2017}.  Moreover,  there is considerable
interest  in understanding the motion of nano-
and micro-particles through biological gels such as mucus which line many of the body cavities,
for example the lungs and the stomach
\cite{Olmsted2001,Lai2007,Lai2009,Kirch2012,Button2012,Ernst2017}.
 %Mucus mainly consists of water and mucins, which are high molecular
 %weight biopolymers that form a heterogeneous biogel
 %[Fig.~\ref{Fig:1}(a) \cite{Kirch2012}] via physical polymer
  % entanglement and additional cross-links between low (electrostatic
  % and hydrophobic) or high (chemical) energy bonds between different
  % polymer segments (see also Fig.~\ref{Fig:1}(b)).  Since the
 %detailed microstructure of mucus, the involved intermolecular forces
 %between mucins, and hence the details of interconnecting cross-links
 %are still under debate (see for example the discussions in
 %Refs.~\cite{Verdugo2012,Ambort2012}), a comprehensive model of mucus
 %does not yet exist.  In the human body mucus plays an important role:
%it lines many organs and traps bacteria and foreign particles.
In order to understand how different
 particles -- proteins, viruses, drugs, and food particles -- cross biological barriers
several experimental studies of
 particle diffusion through mucus have been performed
 \cite{Olmsted2001,Lai2007,Lai2009,Cu2009,Li2009,Kirch2012,Button2012,Kalathi2014,Ernst2017,Ge2017}:
 However, many conventional drugs become trapped in the mucus layer due to steric hindrance or 
 short-range adhesive forces  \cite{Lieleg2011} and therefore it is  of high relevance to
 design new ways of efficiently delivering  drugs based on
 nanoparticles, which could more efficiently cross the mucus barrier \cite{Lai2009}, or move through other biologically complex body fluids such as the
  extracellular matrix \cite{Lieleg2011}.

Simple, diffusion-limited motion in complex fluids is usually too slow for efficient transport.
 A way to overcome this is to use
magnetic particles which can be driven by external
magnetic forces to specific target regions \cite{Kuhn2006,McBain2008}.
 Other ways to drive nano- or micro-particles through  polymeric fluids
 are to use  optical driving forces  \cite{Gutsche2008,Kirch2012} or simply  sedimentation  \cite{Koenderink2004}.
 When a colloid moves in a Newtonian medium of viscosity $\eta$, the expectation is that the
 velocity $V$ follows Stokes formula, $V=F/(6\pi\eta a)$, with $F$ the driving force and $a$ the radius of the particle.
However, Koenderink {\it et al.} showed experimentally that particles driven by
sedimentation in macromolecular xanthan solutions move faster than
expected \cite{Koenderink2004}.  This result has been explained
theoretically by the occurrence of an apparent slip velocity,
experienced by the driven particles because they are surrounded by a uniform
polymer-depleted region \cite{Tuinier2006,Fan2007}.  However, the
density of polymers around a driven particle can be highly
non-uniform, as shown for a colloid dragged through a macromolecular
solution of $\lambda$-DNA with optical tweezers \cite{Gutsche2008}.
Theories have also tried to take into account
the effect of local fluid flow around moving particles, using for
example the concept of dynamic depletion for protein transport in
polymer solutions \cite{Odijk2000}.
Lattice models have recently been used
 to characterize the transport of a driven particle in
 simplified crowded environments \cite{Illien2014}.

%   In Ref.~\cite{Koenderink2004} it has been demonstrated experimentally that this formula is violated, i.e.\ that the
%   transport velocity of sedimenting spheres in polymer solutions is larger than expected.
%   An explanation for this has been given in Refs.~\cite{Fan} which showed that the finite size of the polymers create a polymer-poor layer around the particle which can lead to enhanced particle transport.

 Mesoscale simulations provide a way of investigating the underlying microscopic mechanisms relevant for colloidal motion in polymeric fluids, but currently there are very few 
 simulations of driven colloids in simple polymer solutions~\cite{Ter2005,Gutsche2008}.
 \azz{It has been demonstrated that multiparticle collision dynamics (MPCD) is an efficient method to simulate fluctuating hydrodynamics of colloids (see, e.g.~\cite{Lee2004,Padding2006}) and polymers (see, e.g.~\cite{Malevanets2000,Kikuchi2002,Ripoll2004,Winkler2004,Lee2006,Watari2007,Huang2010}). So far colloid-polymer suspensions  at equilibrium have been studied with MPCD in Refs.~\cite{Li2016,Chen2017,Chen2018   }. Here}
we perform coarse-grained hydrodynamic simulations of driven spheres, ellipsoids and rods moving in a fluid that contains varying concentrations of explicitly modeled, `bead-spring' polymers.
%\ju{I WONDERED WHETHER EXPLICITLY MODELED MIGHT = MOLECULAR DETAIL}

Besides the transport velocities, we determine the fluid flow around the particles and the detailed local polymer properties. Hence we are able to identify  the relevant mechanisms for enhanced colloidal transport in polymer solutions.
%   Also, we do not only model the transport of spheres but also of ellipsoids and rods.
   In Sec.~\ref{sec:meth} we introduce the simulation method, and we present our results in Sec.~\ref{sec:res}.   We summarise our work in Sec.~\ref{sec:con}.

\section{Methods}
\label{sec:meth}
We model the hydrodynamics and fluctuations of the background Newtonian fluids using multiparticle collision dynamics. To simulate polymeric fluids we couple coarse-grained, bead-spring polymers to the solvent. The dynamics of the fluid particles and the colloids is performed using molecular dynamics (MD) simulations.
%Similar approaches to study particle-polymer suspensions but at equilibrium, have been used in Refs.~\cite{Chen2017,Chen2018}. 

\subsection{Multiparticle collision dynamics (MPCD)}
The Newtonian background fluid is simulated using MPCD. This is a coarse-grained solver of the Navier Stokes equations which includes thermal fluctuations \cite{Kapral2008,Gompper2009}. The fluid is modeled by point-like,  effective fluid particles of mass $m$ which perform alternate streaming and collision steps. In the streaming step fluid particles move ballistically for a time $\delta t$ so that their positions $\mathbf{x}_i$ are updated to
\begin{equation}
\mathbf{x}_i (t + \delta t) = \mathbf{x}_i (t) + \mathbf{v}_i(t) \delta t
\end{equation}
where $\mathbf{v}_i$ are their velocities. They are then sorted into cubic cells of length $a_0$ and, in the collision step, all particles in a cell exchange momentum according to
\begin{equation}
\mathbf{v}_i (t + \delta t) = \mathbf{v}_{\xi} (t) + \mathbf{v}_\textnormal{rand}(t) + \mathbf{v}_P(t) + \mathbf{v}_L(t)
\end{equation}
where $\mathbf{v}_{\xi}$ is the instantaneous average velocity in the cell, $\mathbf{v}_\textnormal{rand}$ is a random velocity drawn from a Maxwell-Boltzmann distribution at temperature $T$, and $\mathbf{v}_P $ and  $\mathbf{v}_L $ ensure local linear and angular momentum conservation, respectively  \cite{Gompper2009}.
As basic units in the simulations we chose length $a_0$, mass $m$ and energy $k_BT$.
We use $\delta t = 0.02 \sqrt{ m a_0^2 /k_B T}$ and a  fluid particle number density $n = 10a_0^{-3}$ in order to model viscous flow at  low Reynolds number.
In the absence of polymers the viscosity is then $\eta_0=16.04\sqrt{mk_BT/a_0^4}$.

\begin{table*}
\caption{\label{Tab:1}Simulation parameters.}
\begin{indented}
\item[]\begin{tabular}{@{}|ccccc|ccc|c|c|c|c|}
\br
$\frac{a}{a_0}$ & $\frac{b}{a_0}$ & $\epsilon_1$ & $\epsilon_2$ & description & $\frac{S_x}{a_0}$ &  $\frac{S_y}{a_0}$ &  $\frac{S_z}{a_0}$ & $N_t$ & $\frac{F}{a_{\text{eff}}(k_BT/a_0^2)}$ & $k_b/k_BT$ & symbol/color code \\
\mr
3 & 3 & 1 & 1 & sphere & 48 & 48 & 48 & 70K & 5 & 0 & \includegraphics[height=10pt]{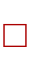} \\
3 & 3 & 1 & 1 & sphere & 48 & 48 & 48  & 70K & 5 & 12 & \includegraphics[height=10pt]{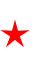} \\
3 & 6 & 1 & 1 & short ellipsoid & 48 & 48 & 60 &  90K & 5 & 0 & \includegraphics[height=10pt]{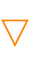} \\
3 & 9 & 1 & 1 & long ellipsoid & 48 & 48 & 120 & 180K & 5 & 0 & \includegraphics[height=10pt]{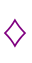} \\
3 & 6 & 0.5 & 1 & short rod &48 & 48 & 60 & 90K & 5 & 0 & \includegraphics[height=10pt]{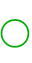} \\
3 & 6 & 0.5 & 1 & short rod &48 & 48 & 60 &  60K & 10 & 0 & \includegraphics[height=10pt]{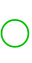} \\
3 & 6 & 0.5 & 1 & short rod &48 & 48 & 60 &  45K & 15 & 0 & \includegraphics[height=10pt]{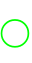} \\
3 & 6 & 0.5 & 1 & short rod &48 & 48 & 60 &  30K & 20 & 0 & \includegraphics[height=10pt]{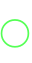} \\
3 & 6 & 0.5 & 1 & short rod &48 & 48 & 60 & 25K & 25 & 0 & \includegraphics[height=10pt]{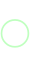} \\
3 & 9 & 0.5 & 1 & long rod & 48 & 48 & 120 &  180K & 5 & 0 & \includegraphics[height=10pt]{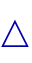} \\
\br
\end{tabular}
\end{indented}
\end{table*}

\subsection{Polymer model}
To simulate polymeric fluids, we add simple bead-spring polymers to the Newtonian background fluid.
Each polymer consists of $N=12$ beads  of diameter $\sigma=a_0$ which are located at positions $\mathbf{r}_i$, $i=1,\dots,N$.
Individual beads are connected by a stiff bond potential
\begin{equation}
  V_\textnormal{bond} = \frac 1 2 k_\textnormal{bond}\sum_{i=2}^{N}(|\Delta \mathbf{r}_i|-l_0)^2
  \label{Eq:Bond}
\end{equation}
%\ju{UNITS? LOOKS LIKE AN EXTRA LENGTH SQUARED ON RHS}
with  $\Delta \mathbf{r}_i = \mathbf{r}_i - \mathbf{r}_{i-1}$, $l_0=a_0$ and $k_\textnormal{bond}=10^5k_BT/a_0^2$. In some cases we include a bending potential
\begin{equation}
V_\textnormal{bend}^P = \frac 1 2 k_\textnormal{b}\sum_{i=3}^{N}\left(\frac{\Delta \mathbf{r}_i \cdot \Delta \mathbf{r}_{i-1}}{|\Delta \mathbf{r}_i| |\Delta \mathbf{r}_{i-1}|} -1\right)^2
\label{Eq:BendP}
\end{equation}
with $k_b=12k_BT$ in order to simulate semi-flexible polymers, but we mainly study systems of flexible polymers where $k_b=0$.

We use a  purely repulsive soft Weeks-Chandler-Anderson (WCA) potential \cite{Weeks1971} between polymer beads,
\begin{equation}
 V_{\textnormal{WCA}}(r) = 
 4\epsilon \left[ {\left( \frac{\sigma^\ast}{r} \right)}^{12} - {\left( \frac{\sigma^\ast}{r}\right)}^{6} \right]  + \epsilon
\label{Eq:WCA}
\end{equation}
for $r < 2^{1/6}\sigma^\ast$ and zero otherwise. Here
 $r$ is the distance between the beads, and we use
$\epsilon=k_BT$ and $\sigma^\ast=\sigma/2^{1/6}$.
We consider fluids at different polymer volume fractions $\rho=\{0.01,0.05,0.1,0.2\}$ with $\rho= N_pN\pi\sigma^3/(6V_d)$ where $N_p$ is the number of polymers in the simulation and $V_d$ the simulation domain volume.
%\ju{DON'T THINK S'S ARE DEFINED YET- COULD SAY V IS THE SIM DOMAIN VOLUME IN UNITS OF AZERO?}
 The polymers are initially randomly distributed in the simulation box,  but are not allowed to overlap with the solid particles.

 \subsection{Colloidal spheres, spheroids and rods}
 We use three different types of solid particles immersed in the polymeric fluids, namely (i) spheres of radius $a$, (ii) ellipsoids with semi-minor axis $a$ and semi-major axis $b$, and (iii) rods of length $2b$ and width $2a$, modeled as a superellipsoid \cite{Barr1981,Balin2017,Zoettl2017}.
 The surface of the superellipsoid at time $t=0$ is given by
%The cell body is modeled by a hard superellipsoid \cite{Barr1981}. Its surface at time $t=0$ is defined by  
\begin{equation}
  \left[\left(\frac{x-x_0}{a}\right)^{\frac{2}{ \epsilon_2}} + \left(\frac{y-y_0}{a}\right)^{\frac{2}{ \epsilon_2}}      \right]^{\frac{\epsilon_2}{ \epsilon_1}} + \left(\frac{z-z_0}{b}\right)^{\frac{2}{ \epsilon_1}} =1
  \label{Eq:super}
\end{equation}
and its  center is located at  ($x_0,y_0,z_0$).
For $\epsilon_1=1$ and  $\epsilon_2=1$ Eq.~(\ref{Eq:super}) reduces to the equation of a conventional ellipsoid with semi-minor axis $a$ and semi-major axis $b$, and if in addition $b=a$, to a sphere.
In order to model rods, we use $\epsilon_1=0.5$ while we keep $\epsilon_2=1$, which means that their cross section is circular.
All the particle dimensions we studied are listed in Table~\ref{Tab:1}.
%\ju{OMIT, REPEATS NEXT PARA  According to Eq.~(\ref{Eq:super}), the colloids are initially oriented in the $z$ direction, but the orientations evolve over time, as described below.}

To drive  the colloids, we apply a force $\mathbf{F}(t)=F \mathbf{n}(t)$ where $\mathbf{n}(t)$ is the instantaneous unit orientation vector pointing along the long axis of the particle. This is initially along the negative $z$ direction (Eq.~(\ref{Eq:super})). The direction of the particle orientation remains constant in time for spheres, but not for the other particle shapes.
In order to obtain approximately the same transport velocity for differently shaped colloids, we normalize the magnitude of the applied force, $F>0$,  by the effective radius $a_{\text{eff}}$ of a particle, defined by $a_{\text{eff}} = \sqrt[3]{3V_p/4\pi}$
where $V_p$ is the volume of the particle.

\begin{figure}
\includegraphics[width=\columnwidth]{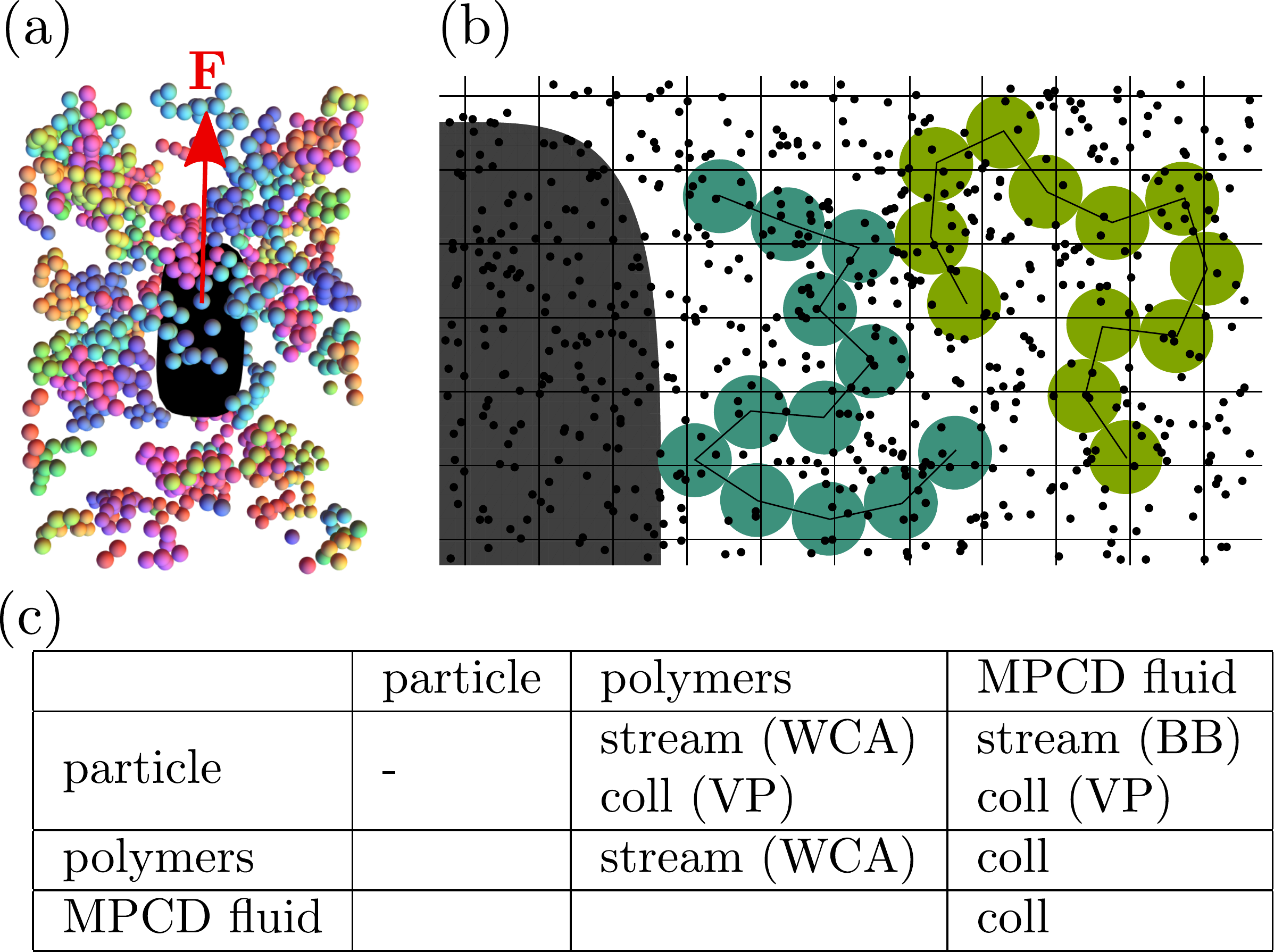}
\caption{\azz{(a) Simulation snapshot around a solid particle (black) driven by a constant force $\mathbf{F}$ through a polymer solution. Different colors of the individual polymers are to aid visualization. (b) 2D sketch of the components involved: driven particle (grey), polymers (light and dark green), MPCD fluid particles (black dots) including virtual particles (VP; black dots in solid particle). The rectangular collision cell grid, which is shifted randomly in every time step \cite{Gompper2009}, is also shown. (c) overview of the interactions between the different components in the streaming  and collision step. Short range forces are realized by  WCA potentials and bounce-back (BB) \cite{Zoettl2018}. }}
\label{Fig:0}
\end{figure}

\subsection{Hybrid MPCD-MD simulations}
To simulate the dynamics of the polymers and the solid particles  in the MPCD fluid we use a hydrid MPCD-MD scheme \cite{Gompper2009,Zoettl2018}. In parallel with  the MPCD streaming step the positions and velocities of the  polymer  beads are updated by determining the forces from the potentials [Eqs.~(\ref{Eq:Bond}) - (\ref{Eq:WCA})] and using a Velocity Verlet algorithm \cite{Allen1989} with time step $\delta t_P = 0.002\sqrt{ m h^2 /k_B T}$. The polymer  beads, which have masses $m_P=10m$, are coupled to the fluid by including them in the collision step \cite{Malevanets2000}. 

The colloid dynamics is also evolved using  a Velocity Verlet algorithm, but with a time step  $\delta t_B = 0.02\sqrt{ m h^2 /k_B T}$. Fluid particles interact with the colloids by applying a bounce back rule, with momentum and angular momentum exchanged accordingly. 
In order to accurately resolve the flow fields near the colloids we use virtual particles inside the colloids, which contribute to the MPCD collision step \cite{Gompper2009}. 
In addition, polymer beads interact with the solid particles via a soft repulsive potential.
\azz{A sketch of the simulated system including an overview of the interactions between the different components involved are shown in Fig.~\ref{Fig:0}.}
%MODIFY PARAGRAPH A BIT COPIED FROM GELSWIMMING PAPER.
%\ju{TRIED A REWRITE - SEPARATING POLYMER DYNAMICS + COLLOID DYNAMICS - AND CALLING THE SOLID PARTICLE A COLLOID OR A BIT CONFUSING}

For all the systems studied we chose the simulation box sizes in the $x$ and $y$ directions to be $S_x=S_y=48$, while varying $S_z$ such that elongated particles minimized self-interaction due to long-range hydrodynamic interactions,  see Table~\ref{Tab:1}.
In addition, we included two \az{hard, impenetrable no-slip} walls, located at $z=\pm S_z/2$, in order to suppress the tendency of the system to self-accelerate  \cite{Hu2015}.
\az{For each simulation a single solid particle is placed inside the simulation box. It is
initialized at position $z_0=+S_z/4$, $x_0=y_0=0$ so that it is not too close to the walls.}
The number of simulation streaming and collision time steps $N_t$ (see  Table~\ref{Tab:1}) is adapted so that particles move  to a final position $z_0 \approx -S_z/4$.
For each system we average over many realizations, i.e.\ between 20 and 65, in order to get good statistics for the measured physical quantities.

\begin{figure}
\includegraphics[width=\columnwidth]{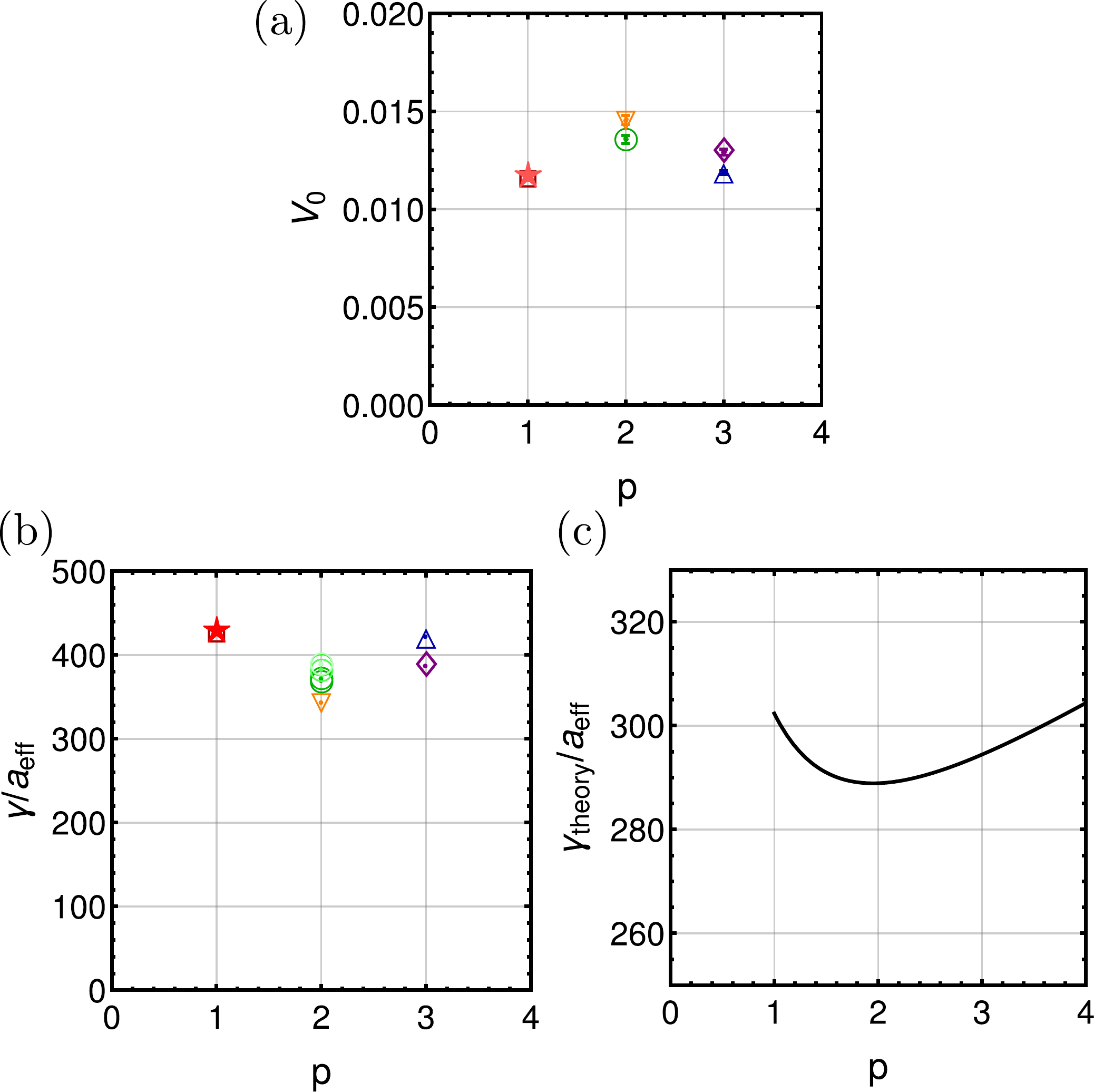}
\caption{(a) Velocity $V_0$ of the differently-shaped colloids  in the absence of polymers as a function of the colloid aspect ratio $p$.
(b) Friction coefficient/effective radius  determined from MPCD simulations for different colloid shapes.
  (c) Theoretical prediction for the friction coefficient/effective radius for prolate ellipsoids.
 Symbol/color code as in Table~\ref{Tab:1}.
  %\\\ju{switch b and c. \\Can we relate the legend to table 1 in some way eg put the colours and descriptons in table 1 - could be 2 column}
 }
\label{Fig:V}
\end{figure}

\section{Results}
\label{sec:res}

\subsection{Velocities: no polymers}
We first determine the time- and ensemble-averaged velocity $V_0$  of the colloids
in the absence of polymers ($\rho=0$) for all the systems listed in Table~\ref{Tab:1}.
In Fig.~\ref{Fig:V}(a) we show results  for different particle geometries, but keeping the driving force per effective radius $F/ a_{\text{eff}}=5k_BT/a_0^2$ constant.
%\ju{ARE DIMENSIONS OK, IS IT BECAUSE ALL LENGTHS ARE NON-DIMENSIONALISED BY A0}
Here $F$ is chosen rather small in order to simulate low Reynolds number flows. We also plot, in  Fig.~\ref{Fig:V}(b), the corresponding friction coefficients, ${\gamma}/{a_{\text{eff}}} = {F}/{(a_{\text{eff}}V_0})$.

As expected, by normalizing the driving force by a colloid's effective radius $a_{\text{eff}}$,  the velocities and friction coefficients of all particles (spheres, ellipsoids and rods) are approximately the same.
There is, however, a small dependence on the particle aspect ratio $p=b/a$ (Fig.~\ref{Fig:V}(a)).
One reason for the dependence of $V$ on $p$ is  that for ellipsoids the translational hydrodynamic friction coefficient, parallel to the long axis of the particles,
divided by the effective radius,   ${\gamma_{\text{theory}}}/{a_{\text{eff}}}$, has a small dependence on the aspect ratio $p$  \cite{KimKarila},
\begin{equation}
  \frac{\gamma_{\text{theory}}}{a_{\text{eff}}} = \frac{16 \pi \eta   p^{-\frac 1 3} }{ \frac{2 p}{1-p^2}+\frac{\left(2 p^2-1\right) \log \left(\frac{\sqrt{p^2-1}+p}{p-\sqrt{p^2-1}}\right)}{\left(p^2-1\right)^{3/2}}}.
  \label{Eq:gamma1}
\end{equation}
This dependence on $p$ is illustrated in Fig.~\ref{Fig:V}(c).

%\ju{NEEDED - GETS IN THE WAY OF THE ARGUMENT? In the absence of polymers the viscosity in the MPCD simulations is $\eta_0=16.04\sqrt{mk_BT/a_0^4}$ for our choice of simulation parameters.}

%From the measurements of $V_0$ we find the friction coefficients in the MPCD simulations using $\frac{\gamma}{a_{\text{eff}}} = \frac{F}{a_{\text{eff}}V_0}$,
%which are plotted in  Fig.~\ref{Fig:V}(c), again as a function of $p$. 
Simulations show the same trend for the dependence of the friction coefficient on $p$ as in the simulations, i.e.\  a minimum for aspect ratio $p=2$.
However, in the simulations, the dependence on $p$ is stronger than the theoretical prediction, and the absolute values of the friction coefficients are larger than the theoretical ones. The most likely reason for the differences is that we do not 
  simulate an infinite domain of fluid, but use periodic boundary conditions for the simulations to be feasible, which can have an impact on the absolute values of the measured friction coefficients \cite{Padding2006}.
More simulations are needed to investigate this further.  However, in the following we will only consider relative, rather than absolute trends in the velocities.

%  \ju{I WOULD HAVE GUESSED FRICTION SMALLER WITH PBC'S} Also our rods are superellipsoids not ellipsoids, and in the simulation the colloids do not always move parallel to their long axis. More simulations are needed to investigate this further.} \ju{HOSTAGE TO FORTUNE HERE - REFS MAY WELL SAY GO  AND DO A FINITE SIZE ANALYSIS}

%DISCUSS ENSKOG FRICTION INFLUENCE?

\begin{figure}
\includegraphics[width=\columnwidth]{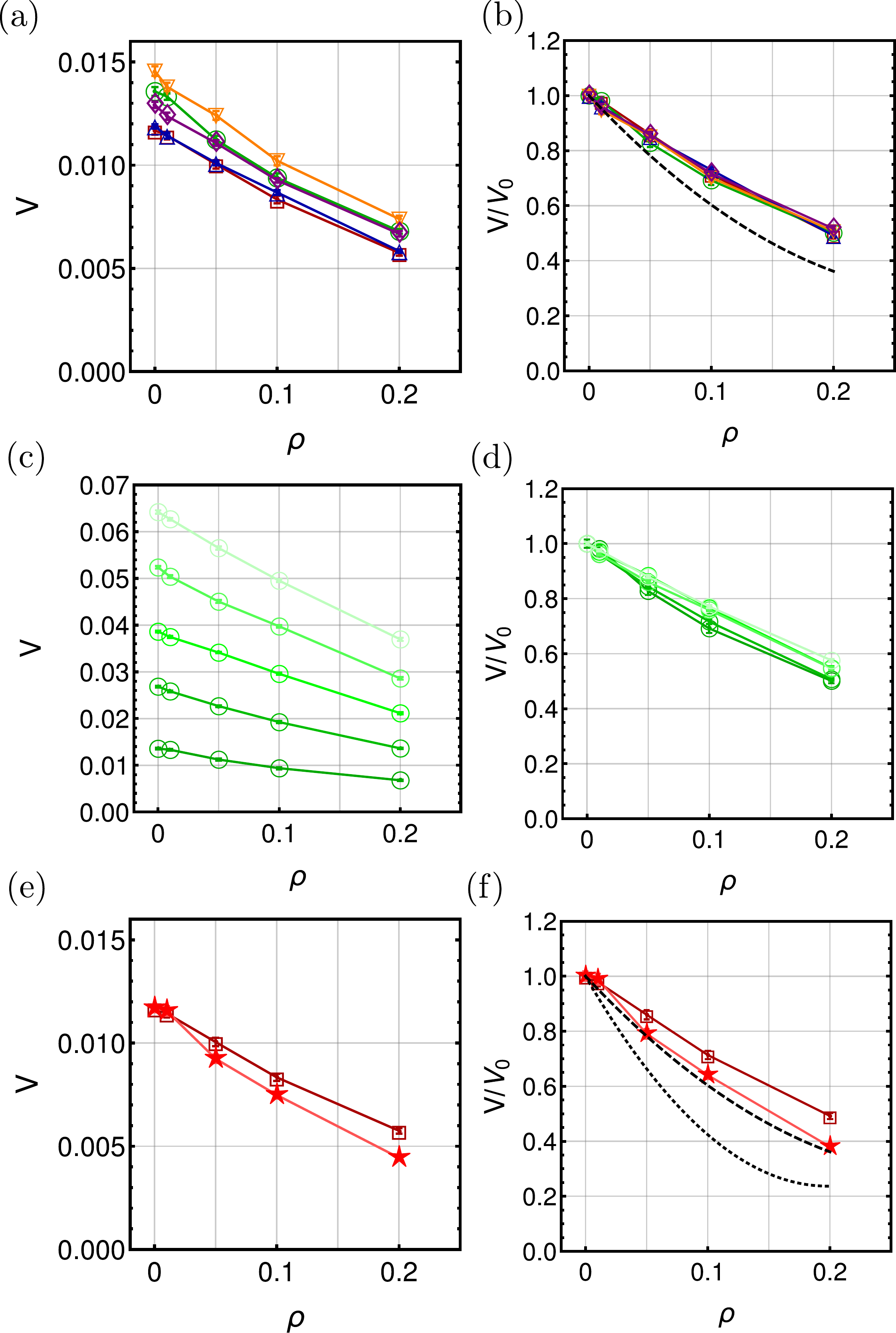}
\caption{
  (a) Time- and ensemble-averaged velocities $V$ as a function of polymer concentration $\rho$ for differently-shaped colloids moving in flexible polymer solutions ($k_b=0$, $N=12$). %\ju{ INSERT COLOR CODE ETC}
  The driving force $F$ is scaled by the effective radius $a_{\text{eff}}$ of the particle: $F/a_{\text{eff}}=5k_BT/a_0^2$.
  (b) Data in (a) with velocities scaled by the zero polymer limit, $V_0$.
  The black dashed line shows the theoretical curve based on simple viscosity scaling.
  (c)  Velocities $V$ for short rods for different driving forces $F/a_{\text{eff}}$ as a function of polymer concentration $\rho$. %\ju{note change - putting in a-eff}
    (d) Data in (c) with velocities scaled by the zero polymer limit, $V_0$.
  (e) Velocities $V$ for spheres for different polymer flexibilities as a function of polymer concentration $\rho$.
  (f) Data in (e) with velocities scaled by the zero polymer limit, $V_0$.
  The black dashed line is the same as in (b). The black dotted line shows the theoretical curve based on simple viscosity scaling for motion in semiflexible polymer solutions.
  %\ju{(a),(b) ... labels will need changing around - easier to change labels than figs\\
  %\ju{In (f) currently two dashed curves which I don't think fits with text and wrong colour
  %  \\box sizes not same yet
  %}
Symbol/color code as in Table~\ref{Tab:1}.
}
\label{Fig:V2}
\end{figure}

\subsection{Velocities: varying the polymer density}

We now add polymers to the fluid, and observe that the velocities $V$ of the colloids decrease with increasing polymer density $\rho$. This is shown in 
Fig.~\ref{Fig:V2}, where again we use a driving force $F/a_{\text{eff}}=5k_BT/a_0^2$.
In Fig.~\ref{Fig:V2}(a) we plot the velocities for spheres, ellipsoids and rods moving in fluids containing flexible polymers ($k_b=0$), which all decay in a similar way.
This is seen more clearly by plotting the rescaled velocities $V/V_0$ against the density $\rho$ in Fig.~\ref{Fig:V2}(b).

Since in Newtonian fluids the friction coefficients increase linearly with the fluid viscosity $\eta$ (see e.g.\ Eq.~(\ref{Eq:gamma1})), the decrease in colloid velocity could simply be due to 
the increase in viscosity with the addition of polymers. If this is indeed the case, the velocity of the particle should scale with the the inverse viscosity $\eta^{-1}$.
In Ref.~\cite{Zoettl2017} we determined the polymer density-dependence of the viscosity  using shear-rheology measurements.
For example, for flexible polymers ($k_b=0$) of length $N=12$, the density-dependence of the inverse viscosity can be fitted to the curve $\eta^{-1}=\eta_0^{-1}(1-4.74\rho+7.75\rho^2)$,
which is plotted as a black dashed line in Fig.~\ref{Fig:V2}(b).
Thus we see that the measured velocities are larger than those predicted by a simple viscosity scaling. We shall show later that the reason for the discrepancy lies in the 
structure of the polymeric fluids.
%, which can not simply be seen as a continous viscous fluid, but its microstructure influences the
%transport velocity of the particles.
%We will eventually find a solution to this discrepancy, as detailed below, which can be found by having a closer look at the fluid velocity fields around the particles, see below.

We also measured the dependence of the particle velocity on the driving force $F$, as shown in Fig.~\ref{Fig:V2}(c).
 $F$ was increased up to $F/a_{\text{eff}}=25k_BT/a_0^2$, where the velocities increase up to $V=0.065\sqrt{k_BT/m}$, corresponding to a Reynolds number $Re=bV/ \nu \approx 0.4$ (using the kinematic viscosity $\nu=1.6\sqrt{k_BT a_0^2/m}$).
As shown in Fig.~\ref{Fig:V2}(d), the scaled velocities $V/V_0$ almost collapse to a single curve showing that the driving force indeed only has a minor effect on the mobility of the particles.
The velocities are, however, slightly higher when using larger driving forces.
As we will see below, this is probably caused by the distribution of polymers around the particles.
%This could be due to finite Reynolds number effects, or to changes in the distribution of polymers around the particle, as discussed in the next section. \ju{RATHER INDECISIVE}
%CHNANGE THISSSSSSSSSSS

Finally we contrast, in Figs.~\ref{Fig:V2}(e) and (f), the velocities of spheres moving in two different polymeric fluids, containing flexible or semiflexible filaments, respectively.
The decay of the velocity $V$ for the sphere moving in the semiflexible polymer solutions is stronger than for flexible polymers.
The main reason for this is that the viscosities of the semiflexible solutions increase more rapidly with density than those of the flexible solutions \cite{Zoettl2017}. However, again, the sphere velocities are larger than expected from a simple viscosity scaling based on the relation $\eta^{-1}=\eta_0^{-1}(1-7.70\rho+19.4\rho^2)$ measured for the semiflexible polymers (see the black dotted curve in  Fig.~\ref{Fig:V2}(f)). %\ju{RED AT THE MOMENT}

Taken together, these results show that the effective mobilities of driven spheres, ellipsoids and rods in a polymeric fluid are larger than those expected from modelling 
the  fluids as a simple continuum viscous medium. We shall now investigate the relation between the structure of the polymeric fluids and the particle velocities.

%We studied the dynamics of driven particles in polymer solutions.
%If not stated otherwise, we always use flexible polymers ($k_B=0$) of length $N=12$.
%The simulation box has 

\begin{figure}
\includegraphics[width=\columnwidth]{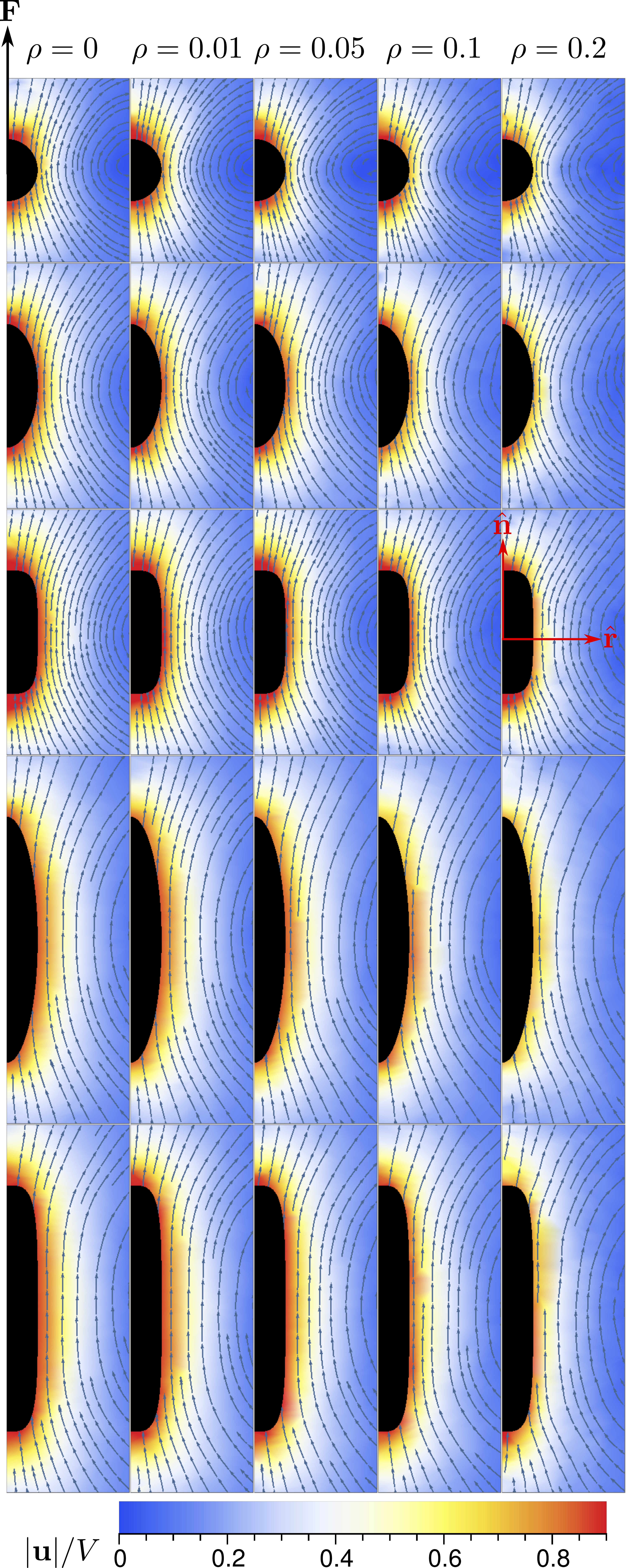}
\caption{Time-, ensemble- and azimuthally averaged flow fields around the different types of colloid (top to bottom) moving in fluids containing flexible polymers of different volume fraction $\rho$.
  Streamlines are shown by blue arrows, and the background color (red to blue) of the plot indicates the strength of the flow field $|\mathbf{u}|$.
  }
\label{Fig:VField}
\end{figure}

\subsection{Flow fields}
To help to understand why the colloids move faster than expected, we measure the velocity fields around them. It is well known that in a Newtonian fluid
a particle driven by a constant force at low Reynolds number creates a Stokeslet flow in the far field, which decays with distance $r$ as $r^{-1}$  \cite{KimKarila}. These long-range flows enable for example sedimenting particle suspensions to interact hydrodynamically and to create large-scale motion \cite{Piazza2014}.
Details of the near-field flow are determined by the shape of the particle.

In Fig.~\ref{Fig:VField} we show the  time-, ensemble- and azimuthally averaged flow fields $\mathbf{u}(\mathbf{r})=u_n\hat{\mathbf{n}}+u_r\hat{\mathbf{r}}$ %\ju{SHOULD THE SUBSCRIPT BE e?}
around single particles driven by a constant force
$\mathbf{F}=F\hat{\mathbf{n}}$ with $F/a_{\text{eff}}=5k_BT/a_0^2$.
The first column displays the flow fields in the absence of polymers, and the other columns the flow fields for solutions containing flexible polymers ($k_b=0$, $N=12$) at different volume fractions $\rho$.
We observe that in all cases the flow fields show a Stokeslet-like behavior away from the particles. In the near field there are strong tangential flows, particularly for the more elongated particles.
In Fig.~\ref{Fig:VField} the strengths of the flow fields $|\mathbf{u}|$ are all normalized to the mean velocities $V$ of the particles.
While their overall shapes do not change significantly with increasing the polymer density, we do observe that the scaled flow field strength
is somewhat suppressed for particles moving in  high-density polymeric fluids (e.g.\ right column with $\rho=0.2$).

\begin{figure}
  \begin{center}
    \includegraphics[width=.85\columnwidth]{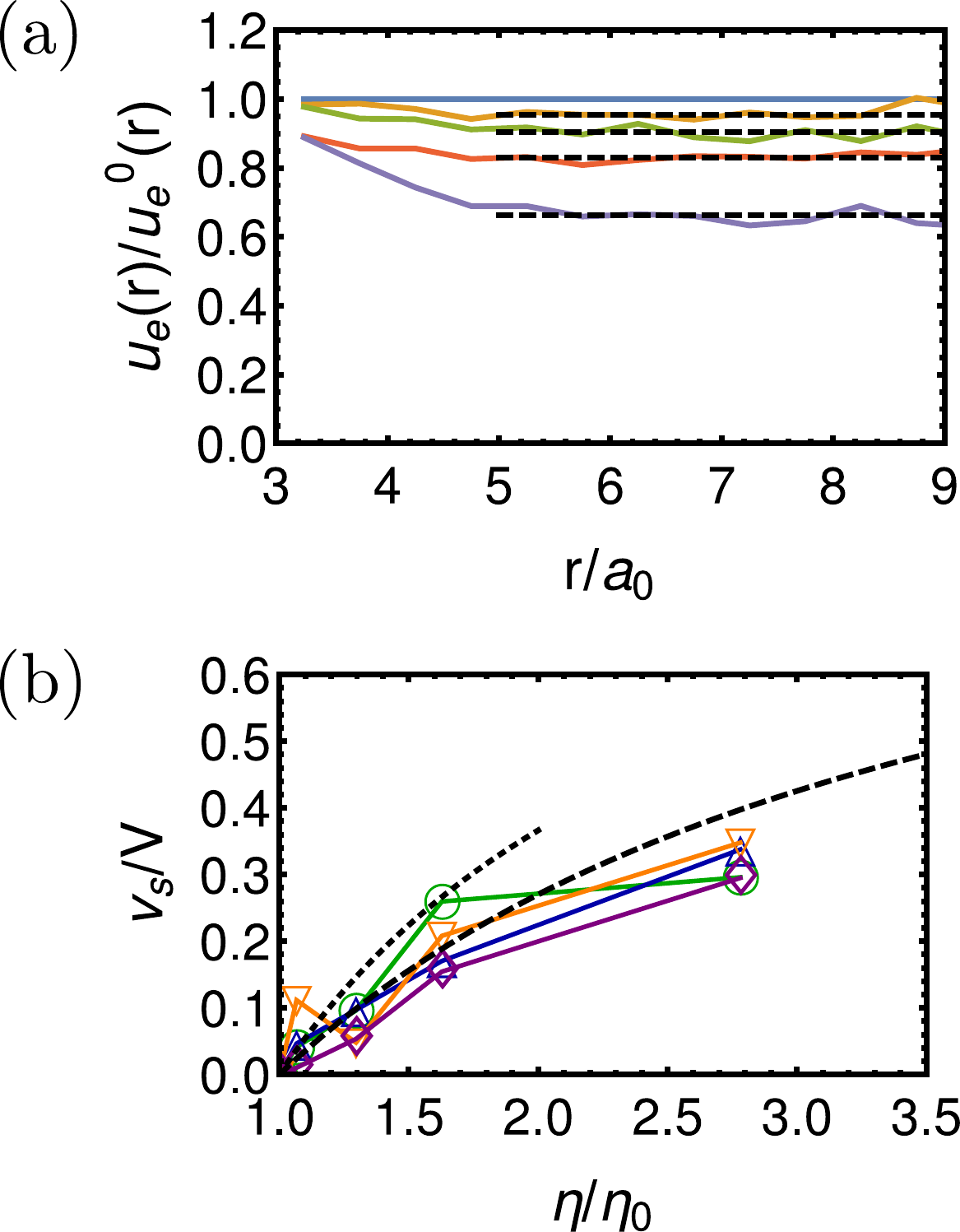}
  \end{center}
  \caption{(a) Decay of the ratio of the non-dimensionalized tangential flow fields $u_e(r)$  and  %\ju{respective? NEEDED}
    no-polymer flow fields $u_e^0(r)$. % for $r>a$ where $a=3a_0$ is the semi-minor axis of the particles.
    The curves, for densities $\rho=0.01$ (orange),  $\rho=0.05$ (green),  $\rho=0.1$ (red) and  $\rho=0.2$ (purple),  level off to constants $I$ (fits shown as black dashed lines).
    (b) Measured apparent slip velocities $v_s/V=1-I$ as a function of the normalized bulk viscosities $\eta/ \eta_0$ of the fluids.
    The black dotted and dashed lines show theoretical predictions from a two-fluid model  with inner fluid-layer thickness $\delta=a_0$ and  $\delta=0.5a_0$, respectively.
    Symbol/color code as in Table~\ref{Tab:1}.
    %\ju{figs would be better larger}
  }
\label{Fig:vs}
\end{figure}

To investigate the differences of the flow fields with and without polymers in more detail,
%\ju{NOT SURE MY VERSION IS RIGHT}
we non-dimensionalize the parallel flow field components \az{$u_n$} with the corresponding particle velocity, defining \az{$u_e=u_n/V$ for the simulation with polymers, and $u_e^0=u_n^0/V_0$ for the no-polymer case.}
%where $u_n^0$ is the parallel flow field component in the absence of polymers.}
%\ju{NOTATION??-SEE ABOVE}
In Fig.~\ref{Fig:vs}(a) we plot the distance-dependent  decays of the ratio $u_e(r)/u_e^0(r)$ measured around the equator of the particle for long rods ($b=9a_0$)
for different polymer densities $\rho$. These results show that the flow fields in the presence of polymers decay more quickly close to the colloid
%(\ju{OMIT BIT CONFUSING WHEN 2 FLUID MODEL USES A0)$r \lesssim 4a_0$)}
than in the polymer-free case. This effect becomes more pronounced with increasing polymer density.
%we compare the non-dimensional, position-dependent velocity components $u_e/V$ and $u_r/V$ with the respective components in the absence of polymers $u_e^0/V_0$ and $u_r^0/V_0$. \ju{REMOVE MENTION OF UR AS DON'T SHOW IT ANYWHERE}
%For example, in Fig.~\ref{Fig:vs}(a) we show the distance-dependent scaled decays of the parallel flow field components $u_e(r)/u_e^0(r)$ measured around the equator of the particle for long rods ($b=9a_0$)
%for different polymer densities $\rho$.
%It can be seen -- in particular for high densities -- that the flow fields in the presence of polymers decay quicker than without the polymers close to the particle ($r \lesssim 4a_0$).
By contrast,  far from from the particle  the ratio $u_e(r)/u_e^0(r)$ levels off to a constant $I$ indicating an $r^{-1}$ scaling in all cases. 
Similar curves are obtained for the other particle shapes although for the spherical particles the flow field ratio does not level off so clearly to a constant. This is because of the finite
simulation box which induces recirculation flows which are strongest for the spherical particles, see  Fig.~\ref{Fig:VField}.

The shape of the flow field can be modelled by introducing an apparent slip velocity $v_s=(1-I)V$  at the surface of the particle  \cite{Lauga2007}. The slip velocities for differently-shaped colloids are shown in~Fig.~\ref{Fig:vs}(b) as a function of the scaled viscosity of the polymer solution.
%\ju{IT SEEMS STRANGE TO SUDDENLY GO FROM TALKING ABOUT DENSITY TO TALKING ABOUT VISCOISTY - BUT I GUESS WE CAN}
%where we plot the extracted apparent slip velocities for different particles but which all show a similar trend.
To understand the reasons for this effective slip around the colloids we next study the properties of the polymers in their vicinity.

\begin{figure}
\includegraphics[width=\columnwidth]{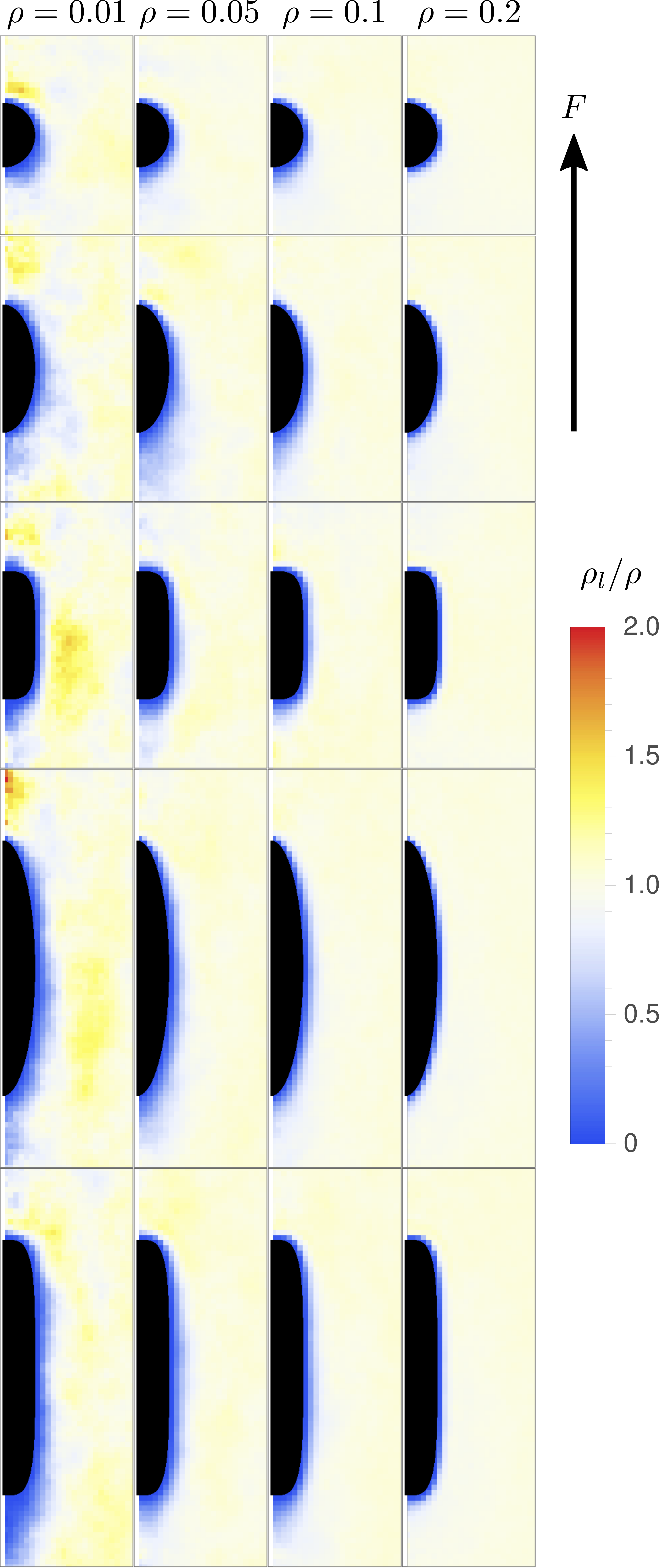}
\caption{
  Local polymer density around different types of particles and at different bulk polymer densities $\rho$.
  The blue layers around the particles indicate  polymer-depleted regions.
  Yellow  indicates polymer-rich regions in front of and next to the particles.
  }
\label{Fig:RhoL1}
\end{figure}

\begin{figure}
\includegraphics[width=\columnwidth]{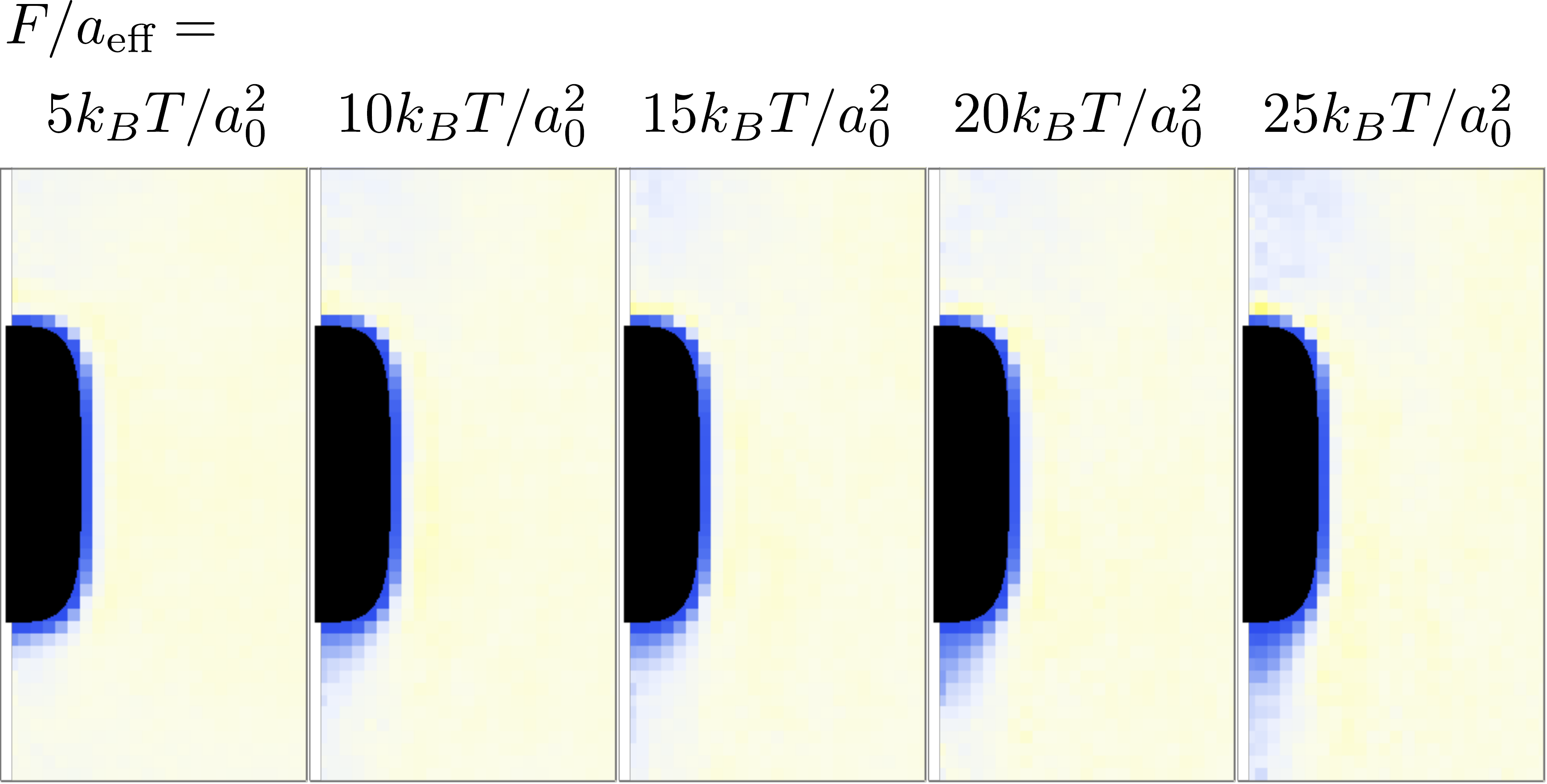}
\caption{Local polymer density for  short rods at different driving forces $F/a_{\text{eff}}$ \az{for $\rho=0.2$}.
  Color code as in Fig.~\ref{Fig:RhoL1}.
  }
\label{Fig:RhoL2}
\end{figure}

\subsection{Local polymer density and the depletion layer}

Explicitly modeling the dynamics of polymers allows us to obtain detailed information about their spatial distribution and their conformations.
In  Fig.~\ref{Fig:RhoL1} we show the time-, ensemble- and azimuthally averaged local polymer densities, $\rho_l$, around the differently-shaped colloids
for polymer solutions of flexible filaments, and using a driving force  $F/a_{\text{eff}}=5k_BT/a_0^2$.

First, it can be seen from Fig.~\ref{Fig:RhoL1} that at high mean polymer density $\rho=0.2$ (right column) the distribution of polymers around all the colloids looks very similar: there is a shallow layer of fluid around the particle (blue region) where the local polymer density is significantly reduced. This layer is of almost constant thickness, and in particular there is no clear front-back asymmetry.

This changes significantly when $\rho$ is reduced. The thickness of the low-polymer-density layer increases, and  a relatively large  polymer-poor region emerges behind the colloids.
For very small $\rho$ (left column) regions of enhanced polymer density in front of and next to the particles also appear.
The reason for this is that polymers are pushed forward by the moving colloid, leading to an enhanced polymer density in front of the particle. This then tends to get pushed sideways
 leading to the polymer-rich regions at the side of the colloid. As the polymers are displaced the moving colloid leaves a polymer-poor area behind, since the polymers need time to diffusively re-enter this region. 
 
 %\ju{MY VERSION NEEDS CHECKING}
For  $F/a_{\text{eff}}=5k_BT/a_0^2$ these effects do not occur at high densities because polymer diffusion and interactions quickly remove any gradients in polymer density. However using higher driving forces, $F$, can lead to significant polymer-poor regions behind the particle even at higher polymer densities, as shown in Fig.~\ref{Fig:RhoL2} \az{for $\rho=0.2$}. The length of the polymer-poor region increases with increasing $F$, as the advective 
 transport of the particles becomes faster than the  diffusion of the polymers. This effect could contribute to the small differences in the transport velocities at different driving forces, see Fig.~\ref{Fig:V2}(d).

%\ju{FIG 7 SHOULD BE BEFORE FIG 6} 
%These regions vanish for larger polymer densities, where the presence of other polymers leads to polymer-polymer interactions leading to a more smooth polymer distribution around the particle.

%This is different at higher polymer densities, where interactions between polymers help them to entropically (??) expand into the polymer-free region fast.
%On the other hand, using higher driving forces $F$ can lead to significant polymer-poor regions behind the particle even at high polymer densities, as shown in Fig.~\ref{Fig:RhoL2}: The higher the driving force, the larger is the polymer-poor area behind the particle.
%This can be seen as a competition between the advective transport of the particles compared to the diffusion of the polymers.
%This effect could contribute to the small differences in the transport velocities at different driving forces, see Fig.~\ref{Fig:V2}(d).

Finally, the reason for the low-polymer-density layer around the particle is the finite size of the polymers.
At low polymer densities polymer depletion layers at surfaces are of the order of the radius of gyration of the polymers  \cite{deGennes1979}.
This is also the case for our polymers ($N=12$) where the radius of gyration is approximately $r_g \approx 2a_0$:  at very low polymer densities (left column of  Fig.~\ref{Fig:RhoL1}), this is the order of the size of the depletion layer.
At higher densities polymer-polymer interactions lead to configurations where the monomers are much more  uniformly distributed around the particle
%\ju{MAYBE JUST STERIC EFFECTS, THEY WANT TO GET AWAY FROM BOTH OTHER POLYMERS AND THE PARTICLE (TO LOWER FREE ENERGY?),}
leading to a  smaller depletion layer thickness.

\begin{figure}
\includegraphics[width=\columnwidth]{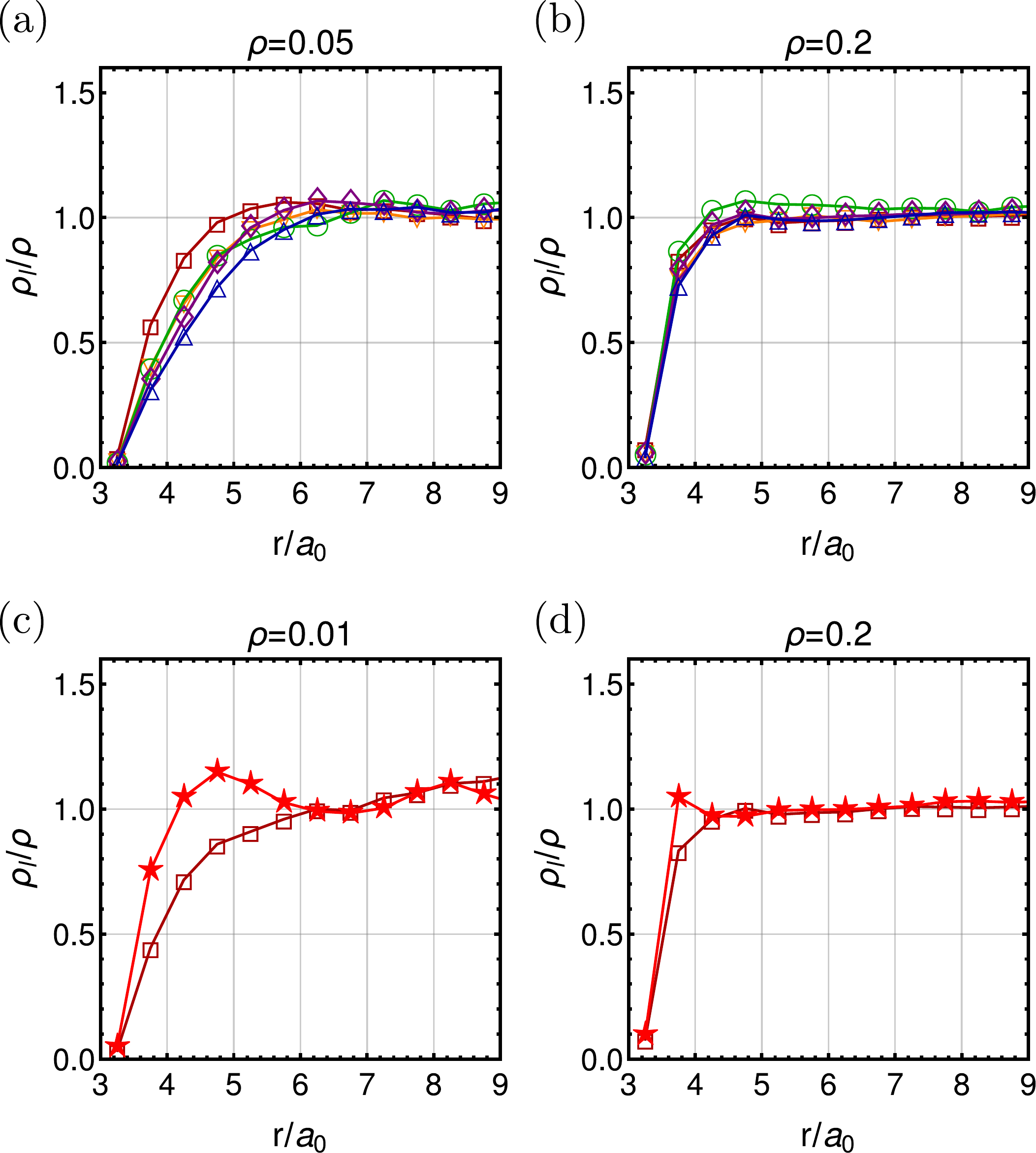}
\caption{Local polymer densities $\rho_l$ normalized by the bulk densities $\rho$. Differently-shaped colloids at polymer densities (a) $\rho=0.05$ and (b) $\rho=0.2$.
  Colloidal spheres in flexible and semiflexible polymer solutions at polymer densities (c) $\rho=0.01$ and (d) $\rho=0.2$.
   Symbol/color code as in Table~\ref{Tab:1}.
  }
\label{Fig:RhoL3}
\end{figure}

A more detailed representation of the polymer distribution around the equator of the colloids is shown in Fig.~\ref{Fig:RhoL3}.
From Fig.~\ref{Fig:RhoL3}(a) we can see that at relatively low bulk polymer density $\rho=0.05$ the polymer density very close to the particle is zero, which means that the local fluid viscosity there is simply $\eta_0$, the fluid viscosity for the no-polymer case.
Then the density of polymers increases gradually, and in a similar way for all particles  considered, to eventually reach the bulk plateau.
The depletion layer thickness, defined as the distance from the particle where the polymer density has dropped to half its value,
$\delta=r(\rho_l/ \rho=0.5) - a $,  is here around $\delta \approx a_0$ and hence for $\rho=0.05$ already smaller than  the radius of gyration.
For higher polymer densities the depletion layer thickness decreases, as shown for the highest density $\rho=0.2$ in Fig.~\ref{Fig:RhoL3}(b).
Here $\delta \approx 0.5a_0$, which is the radius of a monomer.
This is what we expect at high densities, where dense polymer solutions are expected to arrange uniformly due to polymer-polymer interactions, and in our coarse-grained model the monomer size defines the
depletion thickness. Again, this is very similar for all the different colloids considered. %\ju{SO THE LONG THIN ONES TEND TO LIE PARALLEL TO THE SURFACE?}

%\ju{PROBLEM: WE CLAIM THAT BLOB RADIUS = RADIUS OF GYRATION AND THIS DOESN'T SEEM TO FIT THIS ARGUMENT}

We do, however, see an effect of the type of polymer on the depletion layer. % (Fig.~\ref{Fig:RhoL3}).
To show this, we compare results for motion in flexible ($k_b=0$) and semiflexible ($k_b=12k_BT$) polymer solutions.
While for flexible polymers the radius of gyration is a good measure to estimate the depletion layer thickness at low densities, semiflexible polymers create a smaller depletion layer, comparable to the monomer size, see Fig.~\ref{Fig:RhoL3}(c).
%\az{NOT ENTIRELY SURE HOW TO COMMENT HERE, MAYBE LIKE THIS: Note, that locally enhanced polymer density $\rho_l/ \rho > 1$ represents the aforementioned fact that the motion of the colloid push polymeric material to the side. The fact that two peaks for the semiflexible polymers can be identified is most probably an artefact due to the strong density fluctuations and low statistics at low polymer densities.}
At high densities the thickness is again comparable, as shown in Fig.~\ref{Fig:RhoL3}(d), since  it is now determined by the monomer size for both the flexible and semiflexible cases.
%\ju{SEE LAYERING EFFECTS IN THE SEMIFLEXIBLE CASE}

\subsection{Other effects}
So far we have seen that the polymer distribution close to the driven particles is clearly non-uniform.
We also checked other local polymer properties around the particles, such as the local aligning and stretching of the polymers.
However, because we mainly use rather small driving forces, $F/a_{\text{eff}}=5k_BT/a_0^2$, the local polymer properties around the particle are not changed significantly.

\begin{figure}
\includegraphics[width=\columnwidth]{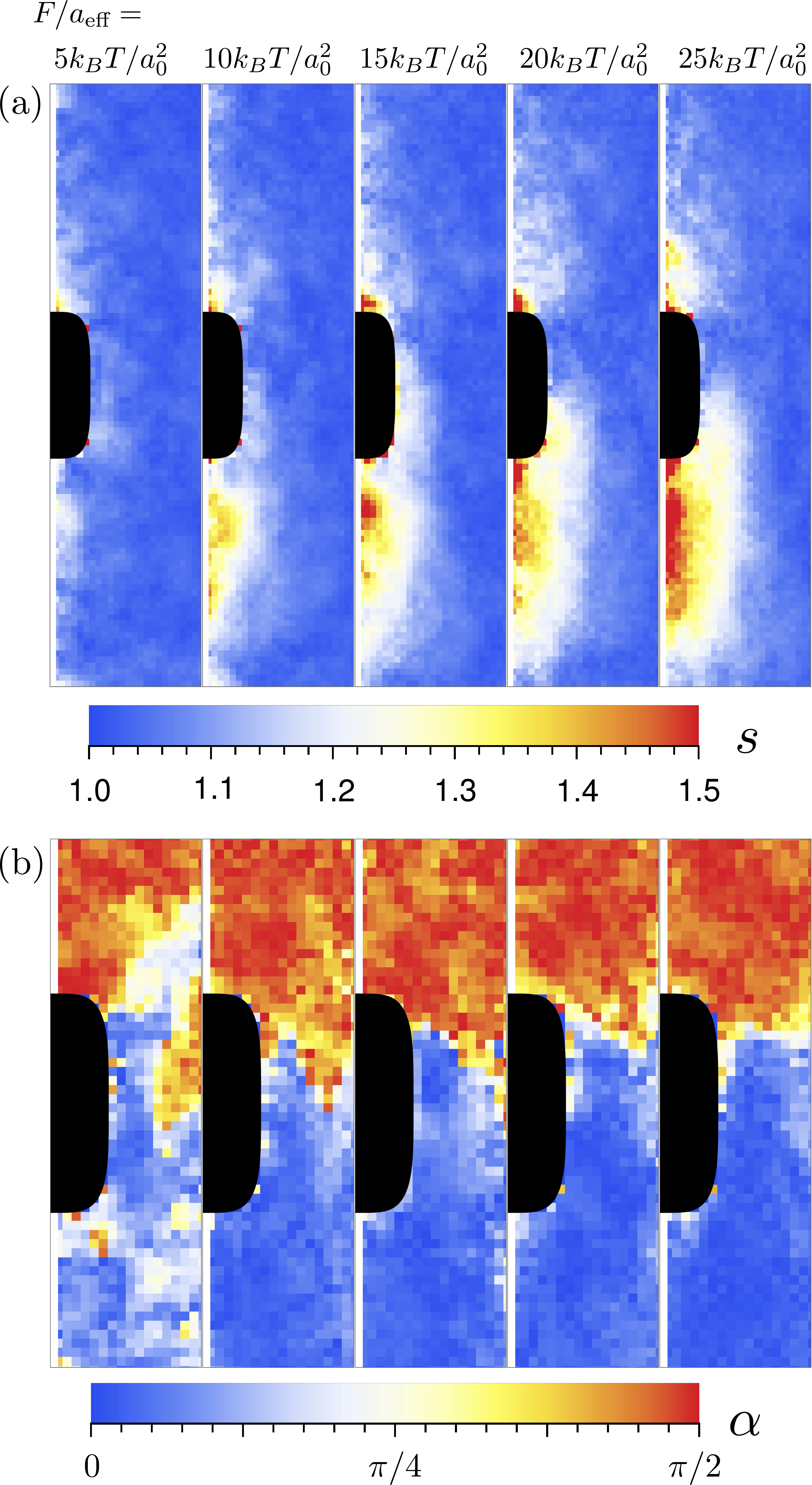}
\caption{\azz{(a) Local effective polymer aspect ratio $s$ [Eq.~(\ref{Eq:s})] for different driving forces $F/a_{\text{eff}}$. (b) Corresponding effective local orientation angle $\alpha$ of the polymers ($\alpha=0$: parallel, $\alpha=\pi/2$: perpendicular to particle orientation.) }
  }
\label{Fig:ETE}
\end{figure}

For larger $F$ we do see some polymer stretching in front of the particle.
\azz{In order to quantify the degree of  stretching of the  polymers, we determine
 % the Eigenvalues $\lambda_1^2 < \lambda_2^2 < \lambda_3^2 $ of
  the
  %time-and ensemble-averaged
  gyration tensors
  %around the particle
  \begin{equation}
    g_{ij} = \frac{1}{N}\sum_{k=1}^N \Delta r_{k,i}\Delta r_{k,j}
  \end{equation}
  for all polymers, at all times and for all different simulation runs,
  where  $k$ is the polymer bead index, and $i$ and $j$ indicate Cartesian components of the relative vector $ \Delta \mathbf{r}_k = \mathbf{r}_k - \mathbf{r}_{c}$ with $\mathbf{r}_{c}$ being the center of mass of the polymer.
  We then compute the local gyration tensors around the particle, averaged over time,  ensembles and azimuthal angles,
  \begin{equation}
    G_{mn}(r,z) = \left\langle A_{mi}(\hat{\mathbf{n}})g_{ij}A_{jn}^T(\hat{\mathbf{n}})   \right\rangle
    \label{Eq:G}
    \end{equation}
  where the transformation matrix $A_{mi}$ depends on the instantaneous particle orientation $\hat{\mathbf{n}}(t)$ and transforms the individual gyration tensors $g_{ij}$  to a coordinate system where the particle orientation $\hat{\mathbf{n}}(t)$ is the first basis vector.
  In Eq.~(\ref{Eq:G}) we averaged over all $g_{ij}$ calculated from polymer beads located within radial distances $r$ and $r+\delta r$, and within longitudinal distances $z$ and $z+\delta z$ from the center of the particle.
  This allows us to not only compute the local eigenvalues $\lambda_1^2(r,z) < \lambda_2^2(r,z) < \lambda_3^2(r,z) $ of $G_{mn}(r,z)$, but also the averaged orientations of the  normalized eigenvectors $\hat{\mathbf{v}}_m(r,z)$ with respect to the particle orientation $\hat{\mathbf{n}}$.
  We measure the strength of the local stretching by the effective polymer aspect ratio $s(r,z)\ge 1$ defined by
  \begin{equation}
    s(r,z) = \frac{\lambda_1}{(\lambda_2+\lambda_3)/2}
    \label{Eq:s}
  \end{equation}
  which is $s \approx 1$ for isotropic conformations and $s \rightarrow \infty$ for highly stretched polymers.

%  Note that, before averaging, we transform the tensor components $G{ij}$ to a coordinate system relative to the orientation of the driven particle in order to compute the orientation of the Eigenvectors $\hat{\mathbf{v}}$ relative to the particle orientation.

  In Fig.~\ref{Fig:ETE}(a)  we plot $s$ for the motion of short rods at polymer density $\rho=0.2$ for different driving forces $F$.
  We can see that polymers get stretched mainly at the back and in front of the particle, and this effect increases with $F$.
  However, the overall effect is rather small for all forces ($s\lesssim 1.5$).
  In order to see how polymers align
with respect to the particle orientation $\hat{\mathbf{n}}$
%more parallel or perpendicular to the particle
we compute the local angle $\alpha$ between  $\hat{\mathbf{n}}$ and the eigenvector $\hat{\mathbf{v}}_1$ corresponding to the largest eigenvalue $\lambda_1^2$ which is in our transformed coordinate system simply $\alpha=\arccos|\hat{v}_{1,x}|$.
  In  Fig.~\ref{Fig:ETE}(b) we can clearly see that in front of the particle the polymers become oriented perpendicular to the particle due to the moving particle  compressing these polymers.
  In contrast, polymers are aligned parallel to the particle at the back because the polymers gain extra space to move in the direction of the particle orientation.
%the local polymer end-to-end distance $\Delta$ compared to the bulk values $\Delta_0$. 
%Note, however, that the effect is in general rather small: even for the highest driving force the polymers are stretched less than  $\sim 20 \%$.
%Moreover, because we use rather short polymers, we do not expect elastic effects to play a major role for the transport of the particles in our system. 
}

\subsection{Two-fluid model}
Our results suggest that it might be possible to represent the fluid around the colloids using a two layer model. This comprises an inner 
layer of thickness $\delta$ which is essentially polymer-free, with viscosity $\eta_0$, and an outer region representing the bulk polymeric fluid, with viscosity $\eta$.
In order to see whether such a two-fluid model can be used to explain the dependence of slip-length on viscosity and the discrepancy between the measured particle velocities and a simple viscosity-scaling approach (Fig.~\ref{Fig:V}(d)), we employ a two-fluid model for spheres discussed in Refs.~\cite{Tuinier2006,Fan2007}.
\az{We use the bulk viscosities $\eta$ and no-polymer viscosity $\eta_0$ obtained from  MPCD simulations \cite{Zoettl2017}.}

For translating spheres with radius $a$ this model can be solved exactly \cite{Tuinier2006,Fan2007}.
It can be shown that the flow fields  decay quickly within the inner layer, and as $r^{-1}$ in the bulk region, and that the
ratio of the flow fields with and without polymers level off to a constant. This agrees with the numerical results in Fig.~\ref{Fig:vs}(a).
Depending on the depletion layer thickness $\delta$ and the viscosity ratio $\eta / \eta_0$ the  apparent slip velocity $v_s$ at the surface of the particle can be calculated as \cite{Fan2007,Zoettl2017}
\begin{equation}
  v_s/V = 1 - \left[ \eta / \eta_0(1+r_s^3/a^3) + r_s^3/a^3 \right]^{-1}
  \label{Eq:vs}
\end{equation}  
where $r_s=a+\delta$.
In Fig.~\ref{Fig:vs}(b) we show the solutions of Eq.~(\ref{Eq:vs}) for $\delta=a_0$ (low-density estimate, black dotted line) and for $\delta=0.5a_0$ (high-density estimate, black dashed line).
We can see that this simple two-layer model indeed captures the measured apparent slip velocities qualitatively, and that even quantitatively the simulation results are in good agreement.
Deviations from the theoretical curves are expected because the crossover between the inner and outer regions is smooth in the simulations, but sharp in the analytic model.

\begin{figure}
  \begin{center}
    \includegraphics[width=.65\columnwidth]{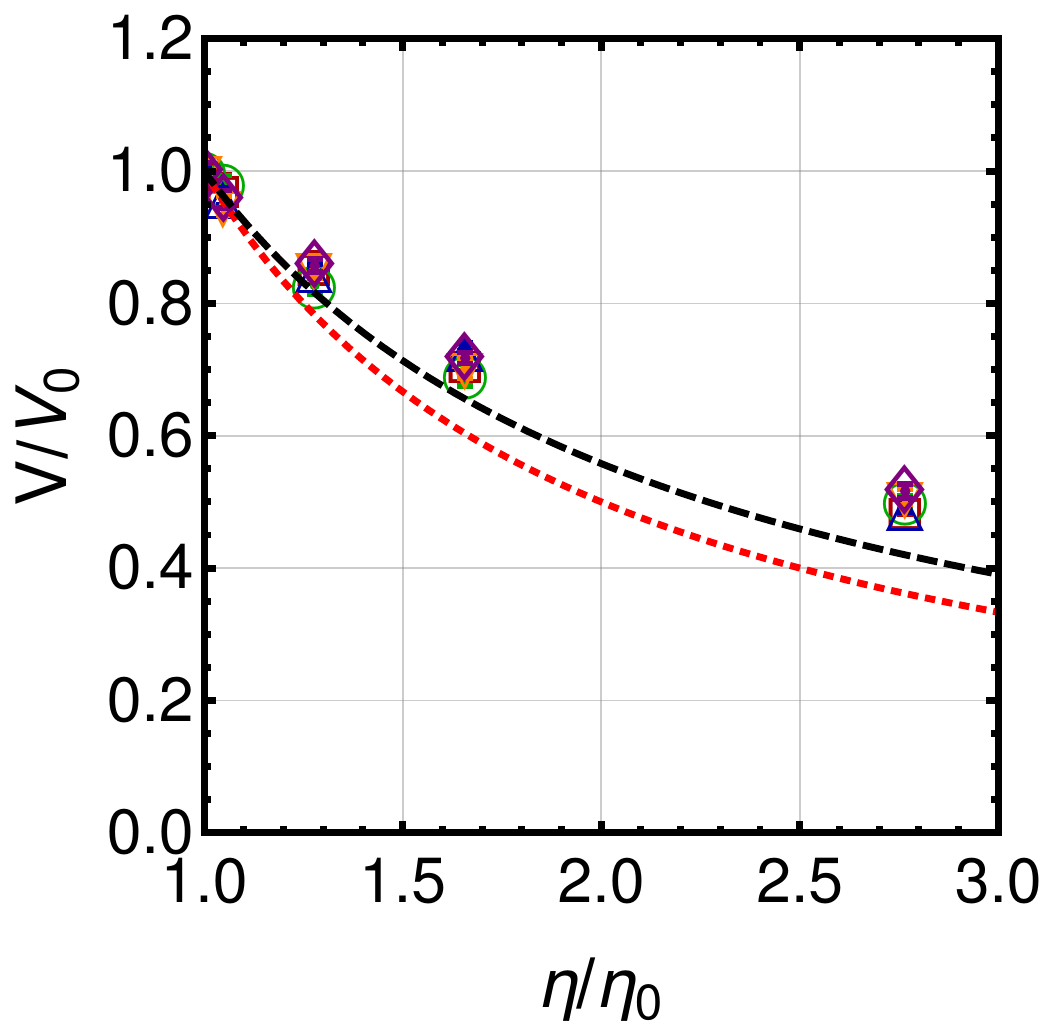}
    \end{center}
  \caption{Velocities, $V$, of the differently-shaped colloids scaled by their respective velocities in a solution with no polymers, $V_0$, as a function of scaled fluid viscosity, $\eta/\eta_0$.
    The black dashed curve shows a theoretical estimate which includes the effect of a finite polymer depletion layer around the particle. The red dotted curve shows the theoretical velocities when the depletion layer is neglected.
    Symbol/color code as in Table~\ref{Tab:1}.
    %\ju{FIG A BIT BIGGER?}
  }
\label{Fig:Vcmp}
\end{figure}

The authors of Refs.\cite{Tuinier2006,Fan2007} also provide a formula for the velocity of the sphere in the presence of the depletion layer.
%Indeed they find that a particle moves faster in the presence of a finite depletion layer compared to the case without a depletion layer.
In Fig.~\ref{Fig:Vcmp} we show how the measured velocities of the spheres, ellipsoids and rods depend on the viscosity of the fluids, and compare the numerical results to the theoretical predictions \cite{Tuinier2006,Fan2007} using a depletion layer thickness $\delta=a_0$ (black dashed line).
Although we do not find perfect quantitative agreement with the theory, the two-fluid model matches our results much better than  simply assuming that the colloid velocity is inversely proportional to the fluid viscosity (red dotted line).

\section{Summary}
\label{sec:con}
We have performed coarse-grained hydrodynamic MPCD simulations of driven spheres, ellipsoids and rods moving in explicitly modeled polymer solutions.
\az{We first determined how the average particle velocities depend on particle shape, polymer densities, driving force and polymer type. We then measured the  flow fields and local polymer density around the particles. Our main finding was that polymer-depleted regions close to the particles are responsible for an apparent tangential slip velocity.
 The thickness of the polymer depleted layer depends on both the density and type of the polymers.
The depletion layer accounts for the measured flow fields and particle velocities, which we capture by a simple model that assumes two layers of different viscosities.}
\azz{We further showed that at sufficiently strong driving forces polymers become stretched and oriented perpendicular to the particle orientation in front of it, and parallel behind it.}

%We demonstrated that the details of the  polymer distributions around the colloids are important in determining their velocities and the fluid flow fields.
Interestingly, there is little difference in the results for spheres, ellipsoids and rods, probably because we have chosen the same semi-minor axis $a$ for all particles. Varying $a$  will be of interest in future work. Moreover, while we performed simulations using simple polymeric fluids, it would be interesting to look for possible effects of shear thinning when using longer polymers, or elasticity if the polymers can create a large number of physical crosslinks.
%\az{In addition, studying the driven motion in }

%\ju{?? THOUGHT IT WAS POLYMER BEAD SIZE AND POLYMER DENSITY which leads to the same ratio between particle size and depletion layer thickness, which governs the speed enhancement.}

%\ju{looking at solvent colloids instead of polymers also interesting}

\section*{Acknowledgments}
%\label{Sec:Ack}
%\addcontentsline{toc}{section}{\nameref{Sec:Ack}}
A.Z.\ acknowledges funding from the European Union's Horizon 2020 research and innovation programme under the Marie Sk\l{}odowska-Curie grant agreement No.\ 653284.
A.Z.\ thanks the Erwin Schr\"odinger Int.\ Institute for Mathematics and Physics for hospitality and financial support through a Junior Research Fellowship.
%TO DO: References, Acknowledgments, text/English    

\section*{References}
\label{Sec:Refs}

\end{document}